\documentclass[aps,preprint,notitlepage]{revtex4-2}

\usepackage{graphicx}
\usepackage{amssymb}
\usepackage{amsmath}
\usepackage{color}
\usepackage{gensymb}
\usepackage{tikz}
\usepackage{float}
\usepackage{datetime2}

\begin{document}
	\raggedbottom
	
	\title{A General Method for Extending Isoscattering Non-Isospectral Quantum Graphs}
	
	\author{Omer Farooq}
	\email{farooq@ifpan.edu.pl}
	\author{Micha{\l} {\L}awniczak}
	\email{lawni@ifpan.edu.pl}
	\author{Leszek Sirko}
	
	\affiliation{Institute of Physics, Polish Academy of Sciences, Aleja Lotnik\'ow 32/46, 02-668 Warszawa, Poland}
	
\date{\today\quad \DTMcurrenttime}

\begin{abstract}
		
The present paper sets out a general method for extending the isoscattering, non-isospectral quantum graphs. The methodology is illustrated by considering two graphs, denoted by $\Gamma_1$ and $\Gamma_2$, which have the same total metric length but different topologies. It is demonstrated that, despite the fact that their scattering matrices are identical, their eigenvalues, denoted by $k_n$, and Euler characteristics differ from one another. The next step in the process is to extend both graphs by affixing the same star graph to each of them.  Utilising the Titchmarsh–Weyl M-function formalism, it is proven that the newly constructed, extended graphs are also isoscattering and non-isospectral.
		
		
\noindent\textbf{Keywords:} quantum graphs, isoscattering, non-isospectral graphs, scattering matrix
		
\end{abstract}
	
	\maketitle	
	
\section{Introduction}

Quantum graphs are metric graphs formed by one-dimensional bonds connected at vertices and equipped with a self-adjoint Laplace operator, $\mathcal{L}(\Gamma)=-\frac{d^2}{dx^2}$, together with appropriate vertex conditions \cite{Berkolaiko2013,Kurasov2024}. In the standard case, the wave function is continuous at each vertex and the sum of the outgoing derivatives vanishes. For finite compact graphs, the spectrum is discrete and non-negative \cite{Gnutzmann2006,Berkolaiko2013,Pluhar2013,Kurasov2024}. Quantum graphs provide an important model of quantum-chaotic wave systems, whose spectral properties can be described within random matrix theory \cite{Gnutzmann2006,Kottos1997,Lawniczak2008,Dietz2017,Rehemanjiang2016,Lu2020,Farooq2024}. Microwave networks constitute convenient experimental realizations of quantum graphs and have been widely used to study their spectral and scattering properties \cite{Hul2004,Lawniczak2010,Lawniczak2019SciRep,Hul2009,Lawniczak2012,Lawniczak2014,Bialous2016,Rehemanjiang2016,Lawniczak2019,Lawniczak2021,Chen2020,Chen2021,Kurasov2025,Lawni2026}.

A fundamental question in spectral theory is how much information about a system can be recovered from spectral or scattering data. For quantum graphs, one may ask whether the topology and metric structure can be uniquely reconstructed from the spectrum or from the scattering matrix \cite{Gutkin2001,Kurasov2005,Band2010,Hul2012}. In general, the answer is negative: distinct graphs may be isospectral, while different open graphs may possess identical scattering matrices \cite{Band2010,Hul2012}. This non-uniqueness is closely related to the classical inverse spectral problem posed by Kac \cite{Kac1966}.
	
Isoscattering graphs $\Gamma_a$ and $\Gamma_b$ have identical scattering matrices, $S_{\Gamma_a}(k)=S_{\Gamma_b}(k)$, where $k$ denotes a wave vector, whereas isospectral graphs have identical eigenvalue spectra, including multiplicities \cite{Band2010,Hul2012}. Thus, isoscattering describes the response of an open graph to waves incident through external scattering channels, while isospectrality concerns the eigenvalue spectrum of the corresponding closed graph.
		
The relation between these two properties has attracted considerable attention. Isospectral quantum graphs with different structures have been constructed using several approaches, and the scattering properties of such systems have also been investigated \cite{Band2010,Hul2012}. More recently, families of isospectral and isoscattering quantum graphs were constructed using symmetries of the graphs and the Titchmarsh--Weyl $M$-function formalism \cite{Kurasov2025}. It was subsequently demonstrated that isoscattering does not necessarily imply isospectrality: quantum graphs may possess identical scattering matrices while the corresponding closed systems have different eigenvalue spectra \cite{Farooq2025,Lawni2025}.
	
In this analysis, we will consider graphs connected at a single vertex, known as a contact vertex, for which the $M$-function is defined as follows:
$M(k) = \partial u(v)/u(v)$,
where $u(v)$ and $\partial u(v)$ denote the values of the function and the sum of the normal derivatives
at the contact vertex $v$.
	
The strength of the M-function formalism lies in its straightforward methodology for analysing the process of graph gluing.   Let us consider a graph $\Gamma$ which is constructed by gluing  two metric graphs $\Gamma_a$ and $\Gamma_b$ at the contact vertices $|\partial \Gamma_a|=|\partial \Gamma_b|$. The M-function for the glued graph is equal to the sum of the M-functions associated with the original parts:
		
		\begin{equation}
			M_{\Gamma}=M_{\Gamma_a}+M_{\Gamma_b}.
			\label{SumM}
		\end{equation}
		
		For any metric graph $\Gamma$ with a single contact vertex, the $S$-matrix and the $M$-function are related as follows \cite{Kurasov2024}:
			\begin{equation} 
			S_\Gamma (k) = \frac{ ik - M_\Gamma (k)}{ik + M_\Gamma (k)}. 
			\label{S}
		\end{equation}
	
	Therefore, any metric graphs with the same M-function exhibit isoscattering behaviour. This is a general property independent of the number of connected subgraphs forming the considered graphs.
	
	In this study, we examine the spectral and scattering properties of two related pairs of quantum graphs. In Case 1, the compact graphs $\Gamma_1$ and $\Gamma_2$   have equal total metric length but different topologies. We obtain their spectra analytically and evaluate the corresponding one-port scattering matrices numerically after coupling each graph to a semi-infinite lead at the contact vertices (CV). We also consider the Euler characteristics of the compact graphs as an additional criterion for testing their non-isospectrality.

	In Case 2, the initial graphs $\Gamma_1$ and $\Gamma_2$ are modified by attaching an identical star graph $\Gamma_S$ at the corresponding CV, giving the extended graphs $\Gamma_{1S}$ and $\Gamma_{2S}$. For these systems, the spectra are determined numerically and the Euler characteristics and scattering matrices are analyzed in the same manner.

\section{Case 1: Initial Graph pair}

\subsection{Graph configuration}

\begin{figure}[H]
	\centering
	\includegraphics[
	width=0.65\linewidth,
	height=0.35\textheight,
	keepaspectratio
	]{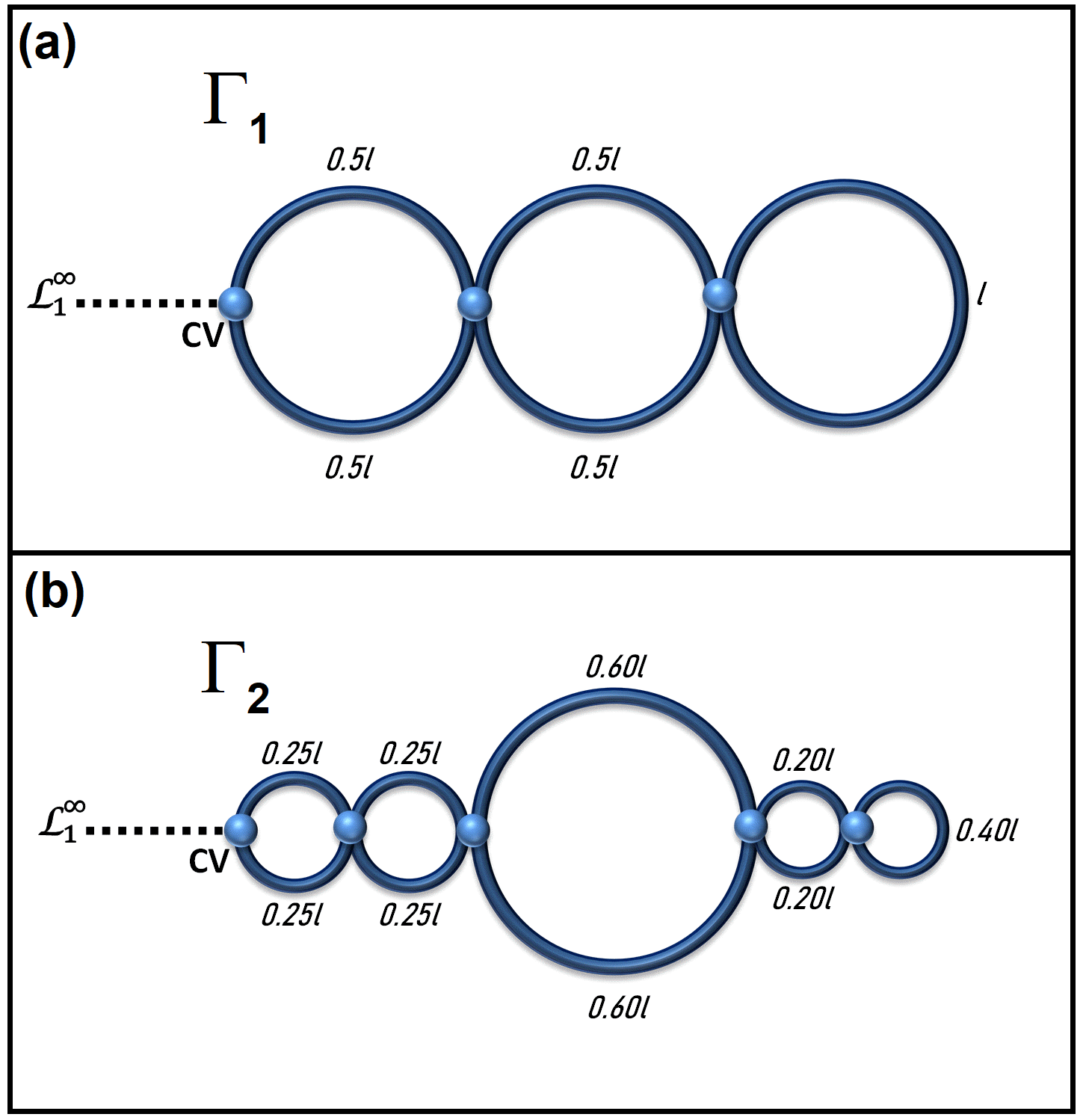}
	\caption{(a) Graphs $\Gamma_1$ and (b) $\Gamma_2$ considered in Case 1. 
		The bond lengths are given in units of $l$. The semi-infinite lead 
		$\mathcal{L}_1^{\infty}$ is attached to the contact vertex CV. 
		The total length of each graph is $L=3l$.}
	\label{fig:case1graphs}
\end{figure}

We consider the quantum graphs $\Gamma_1$ and $\Gamma_2$ shown in Fig.~\ref{fig:case1graphs}. The graphs have different topologies and different distributions of edge lengths, while their total metric lengths are equal. For $\Gamma_1$ the total length is

\begin{equation}
	L_{\Gamma_1}=0.5l+0.5l+0.5l+0.5l+l=3l,
\end{equation}

whereas for $\Gamma_2$

\begin{equation}
	L_{\Gamma_2} =0.25l+0.25l+0.25l+0.25l+0.60l+0.60l+0.20l+0.20l+0.40l=3l.
\end{equation}

Thus,

\begin{equation}
	L_{\Gamma_1}=L_{\Gamma_2}=L.
\end{equation}

In the calculations we set $l=1$ m, and therefore
\begin{equation}
	L_{\Gamma_1}=L_{\Gamma_2}=L=3~\mathrm{m}.
\end{equation}

A semi-infinite lead $\mathcal{L}_1^{\infty}$ is attached to the CV of each graph. We investigate independently the spectra of the corresponding compact graphs and the scattering properties of the graphs coupled to the external lead. Our aim is to show that $\Gamma_1$ and $\Gamma_2$ possess identical scattering matrices despite having different eigenvalue spectra.

\subsection{Analytical spectrum}

The spectra of $\Gamma_1$ and $\Gamma_2$ in Case 1 are determined analytically using the approach described in Ref.~\cite{Pistol2021}. The Laplacian eigenvalues are $\lambda_n=k_n^2$, and in the following the spectra are represented by the corresponding wave numbers $k_n$. For the graphs considered here, the secular equation gives spectral contributions associated with the individual loop lengths and with the total metric length.

For a closed loop of length $\ell_j$, the periodicity condition is

\begin{equation}
	\exp(ik\ell_j)=1,
\end{equation}

which gives

\begin{equation}
	k_n^{(j)}=\frac{2\pi n}{\ell_j},
	\qquad n=1,2,3,\ldots .
	\label{eq:kn}
\end{equation}

The total metric length $L$ gives the additional sequence

\begin{equation}
	k_n^{(L)}=\frac{2\pi n}{L},
	\qquad n=1,2,3,\ldots .
	\label{eq:kn_total}
\end{equation}

The resulting sequences are combined and arranged in increasing order. Coincident values correspond to degenerate levels, with multiplicity determined by the number of coinciding contributions. The zero eigenvalue $k_0=0$ is included once.

For $\Gamma_1$, the three closed loops have lengths

\begin{equation}
	\ell_1=\ell_2=\ell_3=l,
\end{equation}

and the total metric length is

\begin{equation}
	L_{\Gamma_1}=3l.
\end{equation}

For $\Gamma_2$, the five closed-loop lengths are

\begin{equation}
	\ell_1=\ell_2=0.50l,\qquad
	\ell_3=1.20l,\qquad
	\ell_4=\ell_5=0.40l,
\end{equation}

while the total metric length is

\begin{equation}
	L_{\Gamma_2}=3l.
\end{equation}

The spectrum of $\Gamma_2$ is therefore obtained from the sequences associated with these five loop lengths together with the sequence corresponding to the total metric length $3l$. Because the distribution of loop lengths differs from that of $\Gamma_1$, the positions and multiplicities of the resulting eigenvalues are different even though both graphs have the same total metric length.

The overall spectral density can be compared with Weyl's law. For a compact metric graph of total length $L$, the eigenvalue counting function satisfies \cite{Berkolaiko2013,Kurasov2024},
\begin{equation}
	N(k)=\frac{L}{\pi}k+O(1),
	\label{eq:weyl}
\end{equation}
where $N(k)$ denotes the number of eigenvalues up to the wave number $k$, counted with multiplicity. For $L=3\,\mathrm{m}$, \eqref{eq:weyl} gives approximately 20 eigenvalues up to $k=20.96\,\mathrm{m}^{-1}$. The calculated spectra contain 20 eigenvalues for $\Gamma_1$ and 19 eigenvalues for $\Gamma_2$, including multiplicities and the zero eigenvalue. Thus, both spectra are consistent with the same Weyl asymptotics determined by their equal total lengths, although the positions and multiplicities of the individual eigenvalues are different.

The analytically obtained eigenvalues $k_n$ of $\Gamma_1$ and $\Gamma_2$ in the interval $0\leq k\leq20.96$ are listed in Tables~\ref{tab:eigenvalues_gamma1} and \ref{tab:eigenvalues_gamma2}, respectively.
\begin{table}[H]
	\centering
	\caption{Analytically calculated eigenvalues $k_n$ and their multiplicities $m_n$ for $\Gamma_1$.}
	\label{tab:eigenvalues_gamma1}
	
	\normalsize
	\setlength{\tabcolsep}{5pt}
	\renewcommand{\arraystretch}{1.0}
	
	\begin{tabular}{c|ccccccccccc}
		\hline\hline
		$k_n$ [$\mathrm{m}^{-1}$]&
		0 &
		2.09 &
		4.19 &
		6.28 &
		8.38 &
		10.47 &
		12.57 &
		14.66 &
		16.76 &
		18.85 &
		20.94 \\
		
		$m_n$ &
		1 & 1 & 1 & 4 & 1 & 1 & 4 & 1 & 1 & 4 & 1 \\
		
		\hline\hline
	\end{tabular}
\end{table}

\begin{table}[H]
	\centering
	\caption{Analytically calculated eigenvalues $k_n$ and their multiplicities $m_n$ for $\Gamma_2$.}
	\label{tab:eigenvalues_gamma2}
	
	\normalsize
	\setlength{\tabcolsep}{5pt}
	\renewcommand{\arraystretch}{1.0}
	
	\begin{tabular}{c|ccccccccccccc}
		\hline\hline
		$k_n$ [$\mathrm{m}^{-1}$] &
		0 &
		2.09 &
		4.19 &
		5.24 &
		6.28 &
		8.38 &
		10.47 &
		12.57 &
		14.66 &
		15.71 &
		16.76 &
		18.85 &
		20.94 \\
		
		$m_n$ &
		1 & 1 & 1 & 1 & 1 & 1 & 2 & 3 & 1 & 3 & 1 & 1 & 2 \\
		
		\hline\hline
	\end{tabular}
\end{table}
The spectra contain several common eigenvalues, but differ in both the positions and multiplicities of particular levels. The eigenvalues $k=5.24$ $\mathrm{m}^{-1}$ and $15.71$ $\mathrm{m}^{-1}$ occur only for $\Gamma_2$. The eigenvalues
$k=6.28$ $\mathrm{m}^{-1}$, $12.57$ $\mathrm{m}^{-1}$, and $18.85$ $\mathrm{m}^{-1}$ have multiplicity four in $\Gamma_1$, while their multiplicities in $\Gamma_2$ are one, three, and one, respectively. In addition, the eigenvalues $k=10.47$ $\mathrm{m}^{-1}$ and $20.94$ $\mathrm{m}^{-1}$
are nondegenerate in $\Gamma_1$ but have multiplicity two in $\Gamma_2$. Thus, the spectra of the graphs under consideration are different

\begin{equation}
	\sigma(\Gamma_1)\neq\sigma(\Gamma_2),
\end{equation}

showing that $\Gamma_1$ and $\Gamma_2$ are non-isospectral.

\subsection{Euler-characteristic criterion}

The non-isospectrality of $\Gamma_1$ and $\Gamma_2$ can also be seen directly from their topologies. For a finite graph, the Euler characteristic \cite{Kurasov2008Graph,Lawniczak2020Euler,Lawniczak2021Euler} is defined as

\begin{equation}
	\chi(\Gamma)=V-E,
	\label{eq:euler}
\end{equation}

where $V$ and $E$ denote the numbers of vertices and edges of the compact graph, respectively. For metric graphs with standard vertex conditions, the Euler characteristic is determined by the spectrum \cite{Kurasov2024, Kurasov2008Graph,Kurasov2008Schrodinger, Lawni2025,Farooq2023Euler}. Therefore, two isospectral graphs must have the same Euler characteristic.

For the graphs considered in Case 1, $\Gamma_1$ has three vertices and five edges, whereas $\Gamma_2$ has five vertices and nine edges. Hence,

\begin{equation}
	\chi(\Gamma_1)=3-5=-2,
	\qquad
	\chi(\Gamma_2)=5-9=-4.
\end{equation}

Since
\begin{equation}
	\chi(\Gamma_1)\neq\chi(\Gamma_2),
\end{equation}
this topological criterion is consistent with the different spectra listed in Tables~\ref{tab:eigenvalues_gamma1} and \ref{tab:eigenvalues_gamma2}. 

Thus, both the direct comparison of the spectra and the Euler-characteristic criterion lead to the same conclusion: the compact graphs $\Gamma_1$ and $\Gamma_2$ are non-isospectral. We next examine whether this spectral difference is reflected in their one-port scattering properties.
\subsection{Scattering properties}

The scattering properties of $\Gamma_1$ and $\Gamma_2$ are calculated numerically after attaching a semi-infinite lead to the CV. For the one-port configuration considered here, the scattering matrix reduces to the reflection amplitude \cite{Farooq2025,Lawni2025, lawni2023Acta},

\begin{equation}
	S_{11}(k)=|S(k)|e^{-i\varphi(k)}.
\end{equation}

The calculation is performed in the directed-bond representation \cite{Kottos1999, Hul2004}, as also used in pseudo-orbit approaches to quantum graphs \cite{Band2012,Lipovsky2016,Lipovsky2015}. For a graph containing $B$ physical bonds, the internal dynamics is described in a $2B$-dimensional space. The scattering matrix is evaluated according to

\begin{equation}
S(k)=S_4+S_3[I_{2B}-E(\kappa)S_1]^{-1}E(\kappa)S_2,
	\label{eq:scattering}
\end{equation}

where $S_1$ describes scattering between the internal directed bonds, $S_2$ and $S_3$ describe the coupling between the external channel and the internal graph, and $S_4$ represents the direct reflection at the CV. The diagonal propagation matrix $E(\kappa)$ contains the propagation factors associated
with the individual bond lengths $l_j$. In the presence of absorption, these
factors are written as $\exp(i\kappa l_j)$ \cite{Hul2004}.

To include dissipative graphs that can be realized and studied experimentally using microwave networks, absorption is incorporated through a complex wave vector \cite{Hul2004,Lawniczak2014,lawni2023PRE},

\begin{equation}
	\kappa=k+ik_i,
\end{equation}

with

\begin{equation}
	k=\frac{2\pi f}{c},
	\qquad
	k_i=\beta\sqrt{k}.
\end{equation}

Here, $c=299792458$~m/s and the absorption coefficient is
$\beta=0.0090~\mathrm{m}^{-1/2}$. The calculations are performed in the
frequency range from $10$~MHz to $1$~GHz, corresponding approximately to
$0.21\leq k\leq20.96~\mathrm{m}^{-1}$, with a frequency step of
$99.5$~kHz.

For every value of $k$, the modulus $|S(k)|$ and the scattering phase are obtained from the complex reflection amplitude. In the numerical calculations the phase is defined as

\begin{equation}
	\varphi(k)=-\arg S(k).
\end{equation}

\begin{figure}[H]
	\centering
	\includegraphics[
	width=0.70\linewidth,
	height=0.45\textheight,
	keepaspectratio
	]{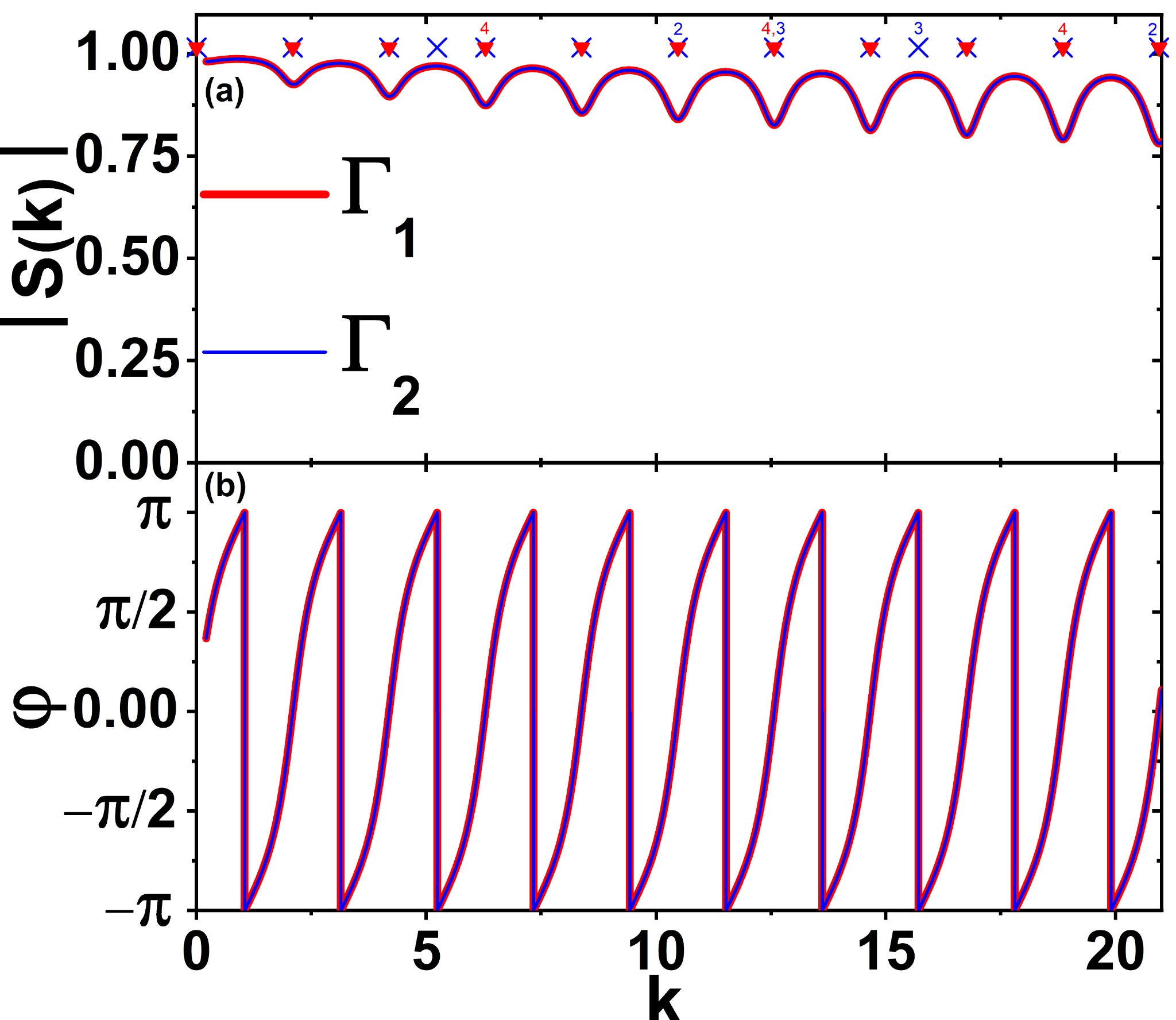}
	\caption{(a) The modulus $|S(k)|$ and (b) the phase $\varphi(k)$  of the 
		scattering matrices calculated for the quantum graphs $\Gamma_1$ 
		(red solid line) and $\Gamma_2$ (blue line). The eigenvalues of 
		$\Gamma_1$ and $\Gamma_2$ are marked in panel (a) by red triangles 
		and blue crosses, respectively. The numbers above the symbols indicate 
		the multiplicities of the corresponding eigenvalues.}
	\label{fig:case1scattering}
\end{figure}

The calculated scattering characteristics of $\Gamma_1$ and $\Gamma_2$ are shown in Fig.~\ref{fig:case1scattering}. Both $|S(k)|$ and $\varphi(k)$ coincide for the two graphs over the investigated range. Therefore,

\begin{equation}
	S_{\Gamma_1}(k)=S_{\Gamma_2}(k),
\end{equation}

showing that $\Gamma_1$ and $\Gamma_2$ are isoscattering \cite{Band2010,Hul2012,Farooq2025,Lawni2025}. Since their eigenvalue spectra are different, the two graphs are isoscattering but non-isospectral.

The result illustrates that the scattering matrix does not necessarily contain the complete spectral information of the corresponding compact graph \cite{Farooq2025,Lawni2025}. The external lead probes the system through the CV. Eigenfunctions that vanish at this vertex may contribute to the spectrum of the compact graph without being visible in the corresponding scattering response. Consequently, different internal spectra may give rise to the same scattering matrix when the graphs are probed through the same CV. This separation between the internal spectral properties and the scattering response motivates the graph extension considered in Case 2.

\section{Case 2: Star Graph Extension}

\subsection{Extended graph configuration}

\begin{figure}[H]
	\centering
	\includegraphics[
	width=0.65\linewidth,
	height=0.35\textheight,
	keepaspectratio
	]{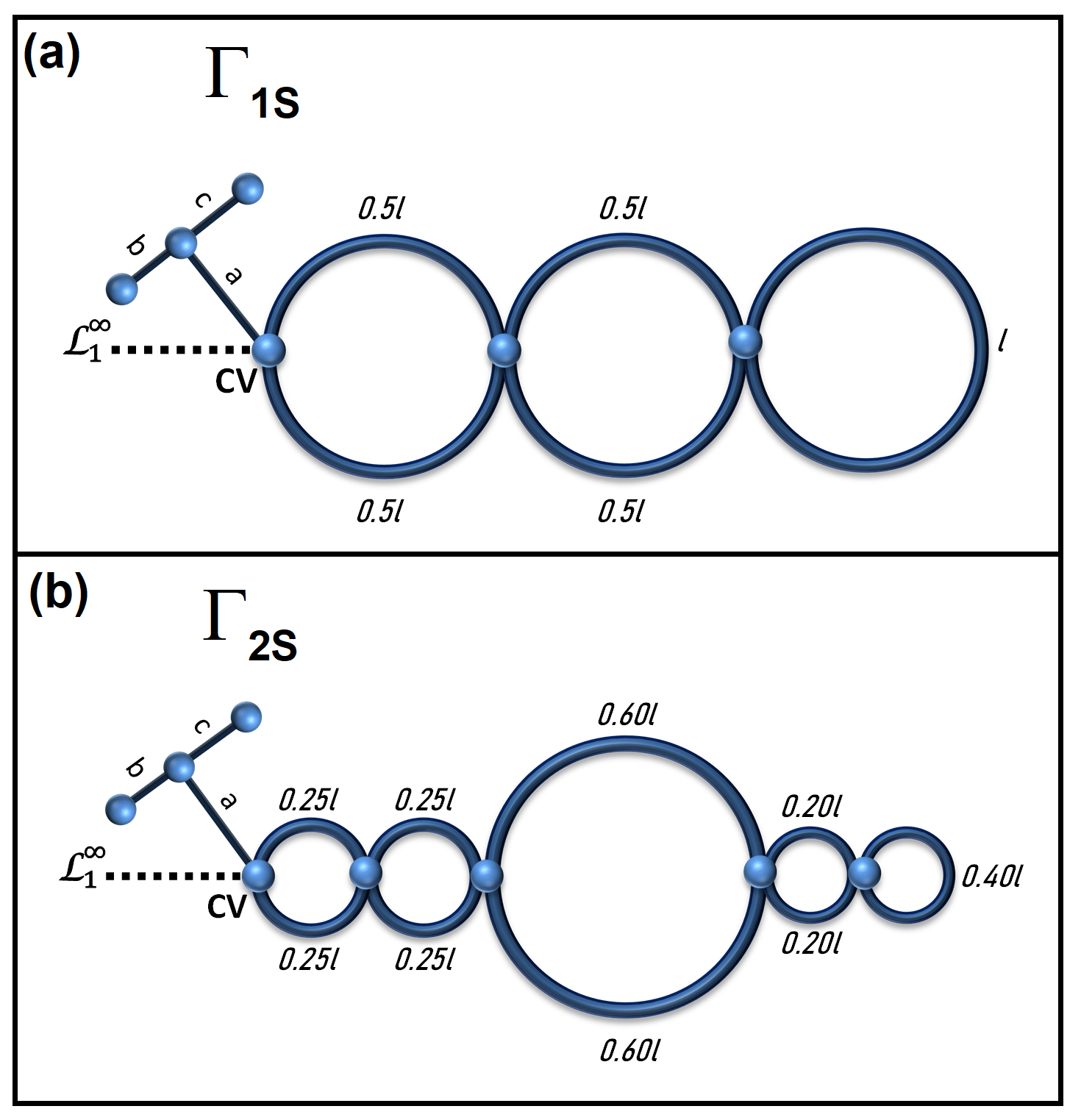}
\caption{(a) Graphs $\Gamma_{1S}$ and (b) $\Gamma_{2S}$ considered 
	in Case 2. The graphs are obtained from $\Gamma_1$ and $\Gamma_2$, 
	respectively, by attaching the same star graph to the contact vertex CV. 
	The bond lengths of the attached star graph are $a=0.50l$ and 
	$b=c=0.25l$.}
	\label{fig:case2graphs}
\end{figure}

In Case 2, the graphs $\Gamma_1$ and $\Gamma_2$ considered in Case 1 are extended by attaching the same star graph $\Gamma_S$  to the corresponding CV. The resulting graphs are denoted by $\Gamma_{1S}$ and $\Gamma_{2S}$, respectively, as shown in Fig.~\ref{fig:case2graphs}. The star graph consists of three bonds of lengths

\begin{equation}
	a=0.50l,\qquad b=c=0.25l,
\end{equation}

giving the total length

\begin{equation}
	L_S=a+b+c=l.
\end{equation}

Since the same star graph is attached to both systems and
$L_{\Gamma_1}=L_{\Gamma_2}=L=3l$, the total lengths of the extended graphs are

\begin{equation}
	L_{\Gamma_{1S}}=L_{\Gamma_{2S}}=L+L_S=4l.
\end{equation}

The original structures of $\Gamma_1$ and $\Gamma_2$ remain unchanged. The purpose of this construction is to investigate whether the isoscattering non-isospectral property of the initial pair is preserved when the same additional graph is attached to both systems.
\subsection{Numerical spectrum}

For the extended graphs $\Gamma_{1S}$ and $\Gamma_{2S}$, the eigenvalues $k_n$ are determined numerically \cite{Kottos1999,Berkolaiko2013}. The internal propagation on the directed bonds is described by the quantum evolution matrix

\begin{equation}
	U(k)=E(k)S_{\mathrm B},
\end{equation}

where $S_{\mathrm B}$ is the bond-scattering matrix and $E(k)$ is the diagonal propagation matrix containing the phase factors associated with the bond lengths. The eigenvalues $k_n$ satisfy the secular equation \cite{Kottos1999,Berkolaiko2013}

\begin{equation}
	\det[I-U(k)]=0.
	\label{eq:secular_case2}
\end{equation}

Using Eq.~(\ref{eq:weyl}) with
$L_{\Gamma_{1S}}=L_{\Gamma_{2S}}=4\,\mathrm{m}$,
approximately 27 eigenvalues are expected up to $k=20.96\,\mathrm{m}^{-1}$. The numerical spectra contain 26 eigenvalues for $\Gamma_{1S}$ and 25 for $\Gamma_{2S}$, including multiplicities and the zero eigenvalue. These small deviations from the leading Weyl estimate are consistent with the $O(1)$ correction in Eq.~(\ref{eq:weyl}). As in Case 1, the common total length determines the leading spectral density but does not require the individual spectra to coincide.

The numerically determined eigenvalues $k_n$ and their multiplicities $m_n$ for $\Gamma_{1S}$ and $\Gamma_{2S}$ are listed in Tables~\ref{tab:eigenvalues_gamma1S} and \ref{tab:eigenvalues_gamma2S}, respectively.

\begin{table}[H]
	\centering
	\caption{Numerically calculated eigenvalues $k_n$ and their multiplicities $m_n$ for $\Gamma_{1S}$.}
	\label{tab:eigenvalues_gamma1S}
	
	\normalsize
	\setlength{\tabcolsep}{8pt}
	\renewcommand{\arraystretch}{1.0}
	
	\begin{tabular}{c|ccccccccc}
		\hline\hline
		$k_n$ $[\mathrm{m}^{-1}]$ & 0 & 1.38 & 2.50 & 4.11 & 5.68 & 6.28 & 6.89 & 8.45 & 10.07 \\
		$m_n$ & 1 & 1 & 1 & 1 & 1 & 4 & 1 & 1 & 1 \\
		\hline
		$k_n$ [$\mathrm{m}^{-1}$] & 11.19 & 12.57 & 13.94 & 15.07 & 16.68 & 18.24 & 18.85 & 19.46 & \\
		$m_n$ & 1 & 4 & 1 & 1 & 1 & 1 & 4 & 1 & \\
		\hline\hline
	\end{tabular}
\end{table}

\begin{table}[H]
	\centering
	\caption{Numerically calculated eigenvalues $k_n$ and their multiplicities $m_n$ for $\Gamma_{2S}$.}
	\label{tab:eigenvalues_gamma2S}
	
	\normalsize
	\setlength{\tabcolsep}{6pt}
	\renewcommand{\arraystretch}{1.0}
	
	\begin{tabular}{c|ccccccccccc}
		\hline\hline
		$k_n$ $[\mathrm{m}^{-1}]$ & 0 & 1.38 & 2.50 & 4.11 & 5.24 & 5.68 & 6.28 & 6.89 & 8.45 & 10.07 & 10.47 \\
		$m_n$ & 1 & 1 & 1 & 1 & 1 & 1 & 1 & 1 & 1 & 1 & 1 \\
		\hline
		$k_n$ [$\mathrm{m}^{-1}$] & 11.19 & 12.57 & 13.94 & 15.07 & 15.71 & 16.68 & 18.24 & 18.85 & 19.46 & 20.94 & \\
		$m_n$ & 1 & 3 & 1 & 1 & 3 & 1 & 1 & 1 & 1 & 1 & \\
		\hline\hline
	\end{tabular}
\end{table}
The spectra of $\Gamma_{1S}$ and $\Gamma_{2S}$ contain several common eigenvalues, but differ in both the positions and multiplicities of particular levels. In particular, $k=5.24$ $\mathrm{m}^{-1}$, $10.47$ $\mathrm{m}^{-1}$, $15.71$ $\mathrm{m}^{-1}$, and $20.94$ $\mathrm{m}^{-1}$ occur only for $\Gamma_{2S}$. The eigenvalues $k=6.28$ $\mathrm{m}^{-1}$ and $18.85$ $\mathrm{m}^{-1}$ have multiplicity four in $\Gamma_{1S}$ but are nondegenerate in $\Gamma_{2S}$, while $k=12.57$ $\mathrm{m}^{-1}$ has multiplicities four and three, respectively. Therefore,

\begin{equation}
	\sigma(\Gamma_{1S})\neq\sigma(\Gamma_{2S}),
\end{equation}

showing that the extended graphs are non-isospectral.
\subsection{Euler-characteristic criterion}

As in Case 1, the non-isospectrality of the extended graphs can also be confirmed independently from their Euler characteristics. The star graph introduces three additional vertices and three additional edges into each graph. Hence,
 the Euler characteristics of the original graphs are preserved:

\begin{equation}
	\chi(\Gamma_{1S})=\chi(\Gamma_1)=-2,
	\qquad
	\chi(\Gamma_{2S})=\chi(\Gamma_2)=-4.
\end{equation}

Therefore,
\begin{equation}
	\chi(\Gamma_{1S})\neq\chi(\Gamma_{2S}),
\end{equation}

which independently confirms that the extended graphs are non-isospectral, in agreement with the numerical spectra listed in Tables~\ref{tab:eigenvalues_gamma1S} and \ref{tab:eigenvalues_gamma2S}.
\subsection{Scattering properties}

\begin{figure}[H]
	\centering
	\includegraphics[
	width=0.70\linewidth,
	height=0.45\textheight,
	keepaspectratio
	]{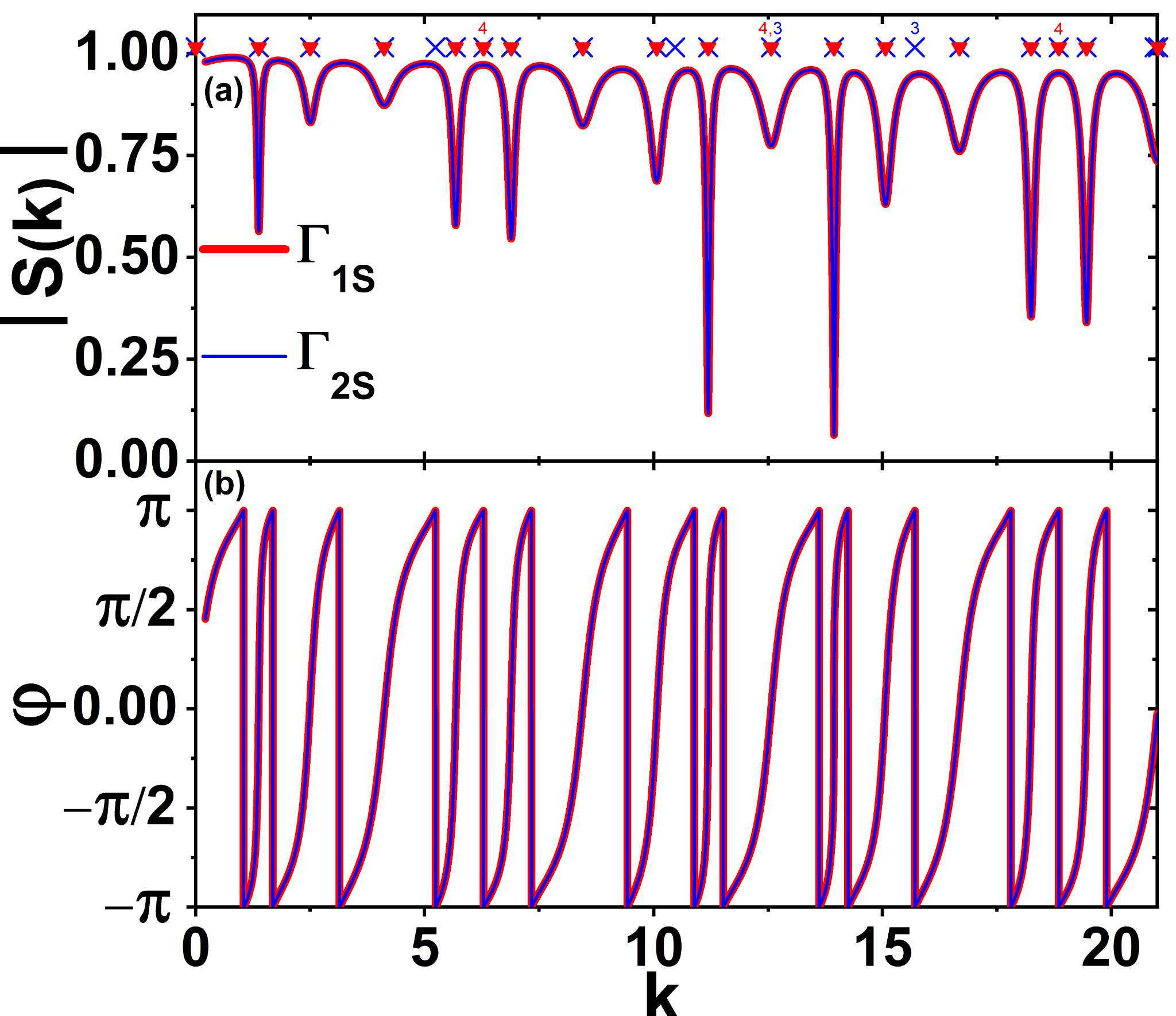}
\caption{(a) The modulus $|S(k)|$ and (b) the phase $\varphi(k)$ of the 
	scattering matrices calculated for the quantum graphs $\Gamma_{1S}$ 
	(red solid line) and $\Gamma_{2S}$ (blue line). The eigenvalues of 
	$\Gamma_{1S}$ and $\Gamma_{2S}$ are marked in panel (a) by red triangles 
	and blue crosses, respectively. The numbers above the symbols indicate 
	the multiplicities of the corresponding eigenvalues.}
	\label{fig:case2scattering}
\end{figure}

The scattering properties of the extended graphs, $\Gamma_{1S}$ and $\Gamma_{2S}$, are investigated using the same numerical procedure employed for the graphs $\Gamma_1$ and $\Gamma_2$ in Case 1.
 The resulting modulus $|S(k)|$ and scattering phase $\varphi(k)$ are shown in Fig.~\ref{fig:case2scattering}.

Both quantities coincide for the two extended graphs over the investigated wave-number range. Therefore,

\begin{equation}
	S_{\Gamma_{1S}}(k)=S_{\Gamma_{2S}}(k),
\end{equation}

showing that $\Gamma_{1S}$ and $\Gamma_{2S}$ are isoscattering. Together with the different numerical spectra and Euler characteristics obtained above, this demonstrates that the extended graphs remain isoscattering and non-isospectral.

The extension of graphs $\Gamma_{1}$ and $\Gamma_{2}$ with graph $\Gamma_S$ modifies the propagation paths and spectral structure of both systems. Nevertheless, the calculated scattering matrices $S_{\Gamma_{1S}}(k)$ and $S_{\Gamma_{2S}}(k)$ remain identical, whereas the spectra of the corresponding compact graphs remain different, $\sigma(\Gamma_{1S}) \neq \sigma(\Gamma_{2S})$.
Thus, for the construction considered here, the initial isoscattering non-isospectral pair of graphs $\Gamma_{1}$ and $\Gamma_{2}$ can serve as a building block for a more complex pair of isoscattering non-isospectral graphs $\Gamma_{1S}$ and $\Gamma_{2S}$.

\section{Conclusions}

A universal approach was developed for expanding the initial isoscattering, non-isospectral quantum graphs into more complex graphs exhibiting equivalent properties. This approach employs the Titchmarsh–Weyl M-function formalism.
In order to demonstrate such properties, the analytical spectra and Euler characteristics of the initial graphs, $\Gamma_1$ and $\Gamma_2$, were calculated. This was done to prove that the graphs are non-isospectral.	In contrast, the calculated one-port scattering matrices demonstrated that they are isoscattering.
Subsequently, the construction of more complex graphs, $\Gamma_{1S}$ and $\Gamma_{2S}$, was achieved by the attachment of a single star graph, $\Gamma_S$, to the corresponding contact vertices of the graphs  $\Gamma_1$ and $\Gamma_2$.
For the extended graphs, $\Gamma_{1S}$ and $\Gamma_{2S}$, the numerical spectra and, consequently, their Euler characteristics are found to be distinct. Concurrently, the calculated scattering matrices remain identical. It can thus be concluded that the spectral distinction and scattering equivalence of the initial pair of graphs, $\Gamma_1$ and $\Gamma_2$, are retained under the extension considered in this study.
The results demonstrate that isoscattering, non-isospectral graphs can be utilised in the construction of more complex graphs without the loss of their general properties. 
Therefore, the construction under consideration provides a direct method of obtaining more complex isoscattering, non-isospectral systems from the simpler ones.

\section*{Acknowledgments}

This research was funded by the Institute of Physics of the Polish Academy of Sciences.


	\end{document}